\documentstyle[preprint,aps]{revtex}

\begin{document}
\title{Reply to Comment `` On large-N expansion''}
\author{Omar Mustafa}
\address{Department of Physics, Eastern Mediterranean University, \\
G. Magosa, North Cyprus, Mersin 10, Turkey\\
email:omar.mustafa@emu.edu.tr}
\date{\today}
\maketitle

\begin{abstract}
Fernandez Comment [1] on our pseudo-perturbative shifted-$\ell $ expansion
technique [2,3] is either unfounded or ambiguous.
\end{abstract}

In his comment [1] on our pseudo-perturbation shifted - $\ell $ expansion
technique (PSLET) [2,3], Fernandez strived to prove that (I) PSLET is just a
version of SLNT, (II) it is not true that PSLET enables one to obtain more
perturbation corrections than SLNT, and (III) it seems that SLNT ( and,
consequently, also PSLET) is divergent.

We explain below why we believe criticisms (I) and (II) to be unfounded and
criticism (III) to be ambiguous.

\-$\bullet ${}{\Large {}{}} Our statement `` the difficulty of calculating
higher-order corrections in SLNT through Rayleigh-Schr\~{o}dinger
perturbation theory (RSPT) results in a loss of accuracy'' is clear and need
not be misleading. We refer to the comprehensive, historical, account in the
work of Imbo et al [4], indicating the actual novelty of SLNT ( which could
handle, via RSPT, only the first four terms of the energy series). Fernandez
and co-workers ( in [6-8] of [1]) have used the hypervirial perturbation
method (HPM) to calculate higher-order corrections in SLNT. Therefore one
would call their method HPM-SLNT, or, at least, {\em Modified} - SLNT (as
they themselves named it) and not SLNT.

$\bullet ${}{\Large {}{}} We did not claim that PSLET is completely
different from SLNT [4] ( c.f. our comment following equation (31) in [3]).
At the top of page 3063 in [5] we commented on the higher accuracy of the
Fernandez HPM-SLNT method (although we had reservations about the
order-dependent shift approach to the Klein-Gordon and Dirac equations).

$\bullet ${} Fernandez derived relations between $a$ and $\beta $, {}$\bar{k}
$ and \-$\bar{l},$ $\cdots $etc. However, that work simply illustrates part
of the message which we tried to deliver to readers, i.e., SLNT is not an
expansion in large-N but, in effect, an expansion in large-$\ell $ ( c.f. C
M Bender et al [6]); hence we preferred the abbreviation PSLET .

$\bullet ${} It is true, of course, that our conclusions in [3] about
numerical accuracy referred to calculations for state wavefunctions with at
most one node. However, the comment by Fernandez that we are unable to apply
our method to wavefunctions with more than one node is unjustified, since we
have in fact given results for such functions in the tables of [2,7,8].
Below we also report PSLET results for the truncated Coulomb potential $%
V(r)=-1/(r+\alpha )$ with $\alpha =10$ for wavefunctions with several nodes.
We show the sum of the first twenty terms of the energy series, $E_{20}$,
and list the corresponding Pad\'{e} approximants. The orders at which the
energy series and Pad\'{e} approximants stabilize are also shown. 
\[
\begin{tabular}{lll}
State & $\ \ \ \ \ \ \ E_{20}$ \ \ \ \ \ \ \ \ 
\begin{tabular}{l}
stability starts \\ 
\ \ \ \ \ from
\end{tabular}
\ \  & \ \ \ \ \ \ \ Pad\'{e} \ \ \ \ \ \ \ 
\begin{tabular}{l}
stability starts \\ 
\ \ \ \ \ from
\end{tabular}
\\ 
4s & \ - 0.011638 \ \ $\ \ \ \ \ \ \ \ E_{12}$ & \ \ - 0.011638 \ \ \ \ \ \
\ \ \ $E[3,3]$ \\ 
6s & \ - 0.006795 \ \ \ \ \ \ \ \ \ \ $E_{12}$ & \ \ - 0.0067958 \ \ \ \ \ \ 
$\ E[7,8]$ \\ 
7s & \ - 0.005443 \ \ \ \ \ \ \ \ \ $\ E_{12}$ & \ \ - 0.0054438 \ \ \ \ \ \
\ $E[7,8]$\  \\ 
9s & \ - 0.003721 \ \ \ \ \ \ \ \ \ \ $E_{13}$ & \ \ - 0.003722 \ \ \ \ \ \ $%
\ \ \ E[7,8]$\  \\ 
11s & \ - 0.002705 \ \ \ \ \ \ \ \ \ \ $E_{14}$ & \ \ - 0.0027068 \ \ \ \ \ $%
\ \ E[8,8]$\ 
\end{tabular}
\]
$\ \ \ \ \ \ \ $

$\bullet ${} Fernandez is unjustified in asserting that PSLET\ is based on
logarithmic perturbation theory (LPT) (c.f. Appendix A in [4] and the
references cited therein on LPT). PSLET is simply an algebraic recursion
method which leads to exactly solvable recursion relations ( based on the
uniqueness of power series representations, c.f. [9]).

$\bullet ${} It is not universally true that HPM-SLNT and consequently PSLET
are divergent. Both techniques are based on asymptotic series expansions and
one would expect to get asymptotically divergent or asymptotically
convergent results (c.f. our analysis in ref.[7,10,11]). To illustrate this
statement with some persuasive evidence we consider the truncated Coulomb
potential with $\alpha =10,$ for wave functions with 10 nodes at $\ell
=1,3,5,15.$%
\[
\begin{tabular}{lllll}
$-E_{M}$ & $\ \ \ \ \ \ \ \ \ell =1$ & $\ \ \ell =3$ & $\ \ \ \ell =5$ & $\
\ \ell =15$ \\ 
$-E_{0}$ & \ \ \ \ 0.00283 \ \  & 0.002198 \ \ \ \  & 0.0017446 \ \ \ \ \  & 
0.00071089 \\ 
$-E_{1}$ & \ \ \ \ 0.00283 & 0.002198 & 0.0017446 & 0.00071089 \\ 
$-E_{2}$ & \ \ \ \ 0.00267 & 0.002118 & 0.0017034 & 0.00070789 \\ 
$-E_{3}$ & \ \ \ \ 0.00256 & 0.002070 & 0.0016809 & 0.00070677 \\ 
$-E_{4}$ & \ \ \ \ 0.00250 & 0.002046 & 0.0016699 & 0.00070636 \\ 
$-E_{5}$ & \ \ \ \ 0.00248 & 0.002034 & 0.0016649 & 0.00070622 \\ 
$-E_{6}$ & \ \ \ \ 0.00247 & 0.002030 & 0.0016629 & 0.00070617 \\ 
$-E_{7}$ & \ \ \ \ 0.00247 & 0.002028 & 0.0016621 & 0.00070616 \\ 
$-E_{8}$ & \ \ \ \ 0.00247 & 0.002028 & 0.0016619 & 0.00070615 \\ 
$\vdots $ & \ \ \ \ 0.00247 & 0.002028 & 0.0016619 & 0.00070615 \\ 
$-E_{20}$ & \ \ \ \ 0.00247 & 0.002028 & 0.0016619 & 0.00070615
\end{tabular}
\]

Obviously, the trends of convergence are very well marked. In general the
energy series of SLNT, HPM-SLNT, and PSLET are oscillatory ( a signal of, at
least, asymptotic convergence) and one would, as a remedy, use an
order-dependent shift ( as in HPM-SLNT) or Pad\'{e} approximants ( as in
PSLET) to obtain results with satisfactory accuracy.

We agree with Fernandez about the unfavorable case ( $\alpha =0.1,$ $\ell
=\nu =0).$ Here the energy series appears to be asymptotically divergent.
However, this should be attributed \ mainly to the nature of the truncated
Coulomb potential and to the irrational value of $\alpha $. One should
notice that this particular potential gives contributions to the
higher-order corrections of the energy series through its non-vanishing
higher-order derivatives. This will lead to accumulated rounding-off errors
which, in turn, can yield unreliable results from the higher-order
corrections.

We believe that the points made above have satisfactorly answered the
criticisms (I) to (III) of Fernandez.

\end{document}